\documentclass[
reprint,
floatfix,
superscriptaddress,
amsmath,
amssymb,
aps,
prl
]{revtex4-2}

\usepackage{graphicx}
\usepackage[utf8]{inputenc}
\usepackage[T1]{fontenc}
\usepackage{siunitx}
\usepackage[
    colorlinks=true,
    citecolor=blue,
    linkcolor=blue,
    urlcolor=blue
]{hyperref}
\newcommand{\nmrad}[1]{\ensuremath{#1\,\mathrm{nm}\text{-}\mathrm{rad}}}

\begin{document}

\title{Transverse Confinement in Cascaded Inverse-Cherenkov Acceleration via Phase-Trajectory Control and Injection Matching}

\author{Yuanhao Liu}
\affiliation{Shanghai Institute of Applied Physics, Chinese Academy of Sciences, Shanghai 201800, China}
\affiliation{University of Chinese Academy of Sciences, Beijing 100049, China}

\author{Nanshun Huang}
\affiliation{Shanghai Advanced Research Institute, Chinese Academy of Sciences, Shanghai 200120, China}

\author{Xionghao Yuan}
\affiliation{Shanghai Institute of Applied Physics, Chinese Academy of Sciences, Shanghai 201800, China}
\affiliation{University of Chinese Academy of Sciences, Beijing 100049, China}

\author{Xingheng Wang}
\affiliation{Shanghai Advanced Research Institute, Chinese Academy of Sciences, Shanghai 200120, China}

\author{Juhao Wu}
\email{wujuhao2026@sari.ac.cn}
\affiliation{Shanghai Advanced Research Institute, Chinese Academy of Sciences, Shanghai 200120, China}

\author{Haixiao Deng}
\email{denghx@sari.ac.cn}
\affiliation{Shanghai Advanced Research Institute, Chinese Academy of Sciences, Shanghai 200120, China}

\date{\today}

\begin{abstract}
Sub-relativistic electrons in micron-scale dielectric laser accelerators face coupled challenges of longitudinal dephasing and transverse defocusing, often imposing stringent requirements on injector emittance. Here, we demonstrate a combined strategy of phase-trajectory control and injection matching for sustained, confined acceleration. By coordinating the refractive-index and section-length sequences with the injected beam's Twiss parameters, we control accumulated phase slippage while maintaining transverse confinement. In a seven-section inverse-Cherenkov accelerator simulation, a 100-keV electron bunch with an initial normalized transverse emittance of \nmrad{1} and a relative kinetic-energy spread of \SI{0.1}{\percent} reaches \SI{1154.7}{keV} over \SI{1.545}{mm}, with a transverse rms beam size below \SI{1}{\micro\meter} throughout the accelerating channels. The exit normalized emittance is \nmrad{1.92}, and the relative RMS kinetic-energy spread is \SI{0.6}{\percent}, with no particle loss under the specified tracking-acceptance criterion. The results demonstrate that coordinated structural sequencing and injection matching can relax injector-emittance requirements by up to approximately two orders of magnitude, offering a route toward compact dielectric accelerators for ultrafast electron diffraction, microscopy, and chip-scale coherent radiation.
\end{abstract}

\maketitle

Particle accelerators play essential roles in medicine, industry, and scientific research~\cite{1}; however, conventional radio-frequency accelerators are usually large and costly. Compact alternatives are therefore of considerable interest and include advanced radio-frequency structures, plasma wakefield accelerators, and terahertz (THz)-driven accelerators. Among these approaches, dielectric laser accelerators (DLAs) exploit optical or THz fields in dielectric microstructures to achieve high accelerating gradients, offering potential for on-chip integration~\cite{7}. Two representative acceleration mechanisms used in DLAs are inverse Smith--Purcell (ISP) acceleration and inverse Cherenkov (ICR) acceleration.

ISP-DLAs employ spatial harmonics excited in periodic dielectric gratings to accelerate electrons~\cite{6,10,11,8}. Experiments have demonstrated accelerating gradients of \SI{25}{MeV\per\meter} for \SI{28.0}{keV} electrons and more than \SI{250}{MeV\per\meter} for relativistic electrons~\cite{8,9,12}. Subsequent advances in phase-space control and alternating-phase focusing have enabled coherent energy gains of \SI{12.3}{keV} and improved beam confinement~\cite{13,14,15,16,17}. However, maintaining phase synchronization during acceleration remains challenging for grating-based structures. In contrast, ICR-DLAs utilize evanescent fields at dielectric interfaces, with the phase velocity controlled by the refractive index and incidence angle~\cite{18,20,21,19}. Compared with grating-based ISP-DLAs, ICR-DLAs provide extended synchronous acceleration without requiring longitudinally periodic gratings or an external pulse-front-tilt system, offering advantages in structural simplicity and optical-system integration~\cite{19,22,23}. An energy gain of \SI{0.8}{keV} and an accelerating gradient of $19\pm2$~MeV/m have been achieved experimentally~\cite{18}. Numerical studies have further explored prism-based, multichannel, and cascaded ICR-DLAs~\cite{24,22,25,26,27,28}, including an alternating-phase focusing design with an average accelerating gradient of \SI{220.8}{MeV\per\meter}~\cite{23}.

Despite these advances in DLA research, sustained acceleration of sub-relativistic electrons remains limited by longitudinal phase slippage and transverse de-focusing, which unavoidably cause beam expansion and particle loss~\cite{23,28,29}. Although alternating-phase focusing and dynamic-phase control have shown promising results~\cite{23,28,30}, current numerical studies still employ nearly cold electron beams. For example, an alternating-phase focusing study adopted an initial normalized transverse emittance of 20~pm-rad~\cite{23}, whereas a more recent study of dynamic phase-driven cascade focusing assumed an initial normalized transverse emittance of 10~pm-rad~\cite{28}. To address these challenges, we propose a phase-trajectory control and injection-matching strategy for bunches with finite emittance and nonzero energy spread. The refractive index $n_i$ controls the local phase-velocity mismatch, while the section length $L_i$ regulates accumulated phase slippage. Combined with injection Twiss-parameter matching optimization, this approach aims to achieve sustained acceleration and transverse confinement. A seven-section symmetric dual-prism ICR-DLA serves as a numerical demonstration, with three-dimensional particle tracking of a bunch having an initial normalized transverse emittance of \nmrad{1} and a relative rms kinetic-energy spread of \SI{0.1}{\percent}.

\begin{figure}[!t]
\centering
\includegraphics[width=0.98\columnwidth]{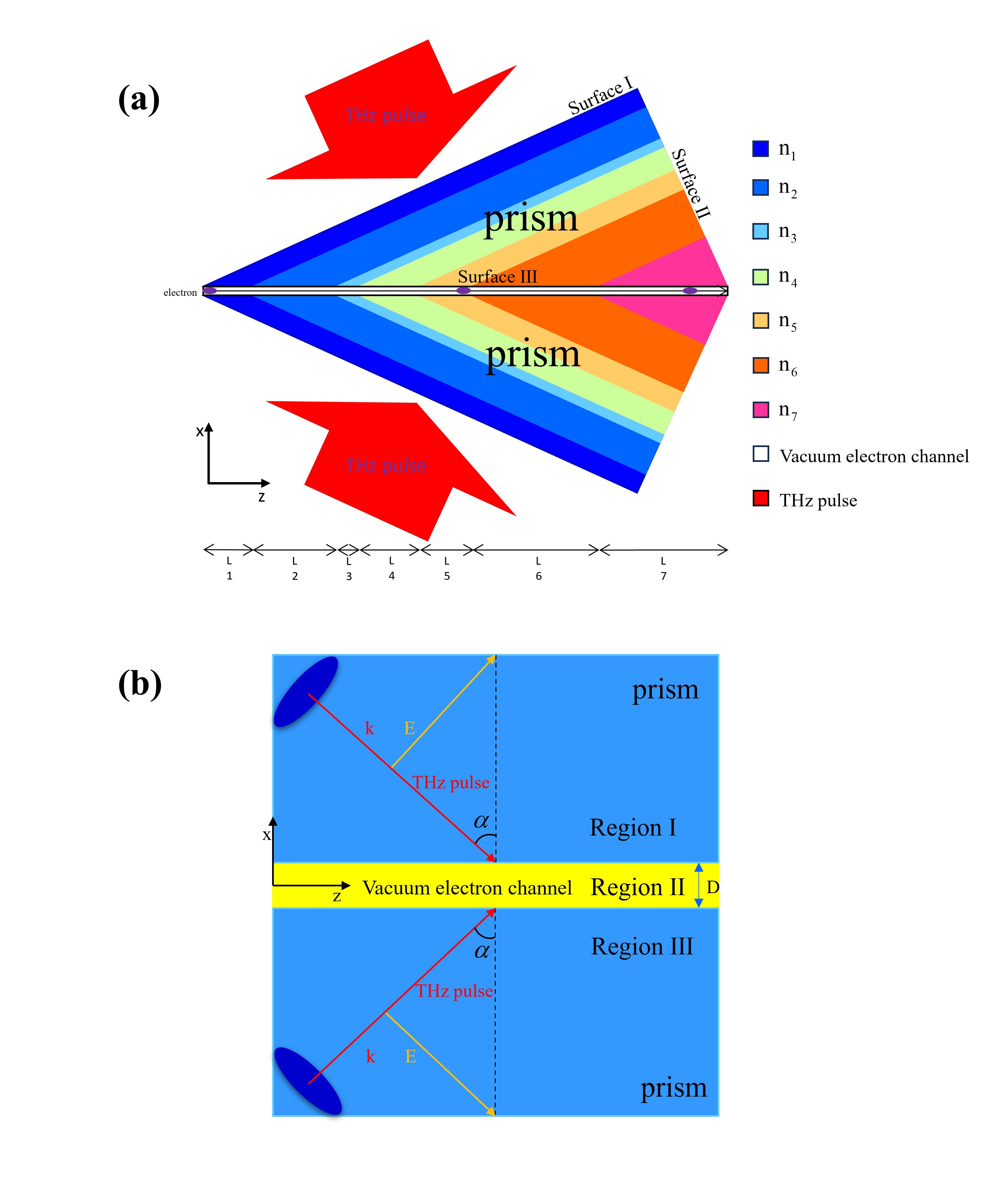}
\caption{Symmetric dual-prism ICR-DLA. (a) Seven sections with indices $n_i$ and lengths $L_i$ are driven by equal, in-phase THz pulses. (b) Local $x$--$z$ geometry; $\alpha$ is measured from the interface normal and $D$ is the vacuum-channel width.}
\label{fig:structure}
\end{figure}

We use a local carrier-frequency model to interpret the applied near-single-cycle THz waveform. For a homogeneous section Fig.~\ref{fig:structure},
\begin{align}
k_z&=nk_0\sin\alpha, &
\kappa&=k_0\sqrt{n^2\sin^2\alpha-1},\nonumber\\
v_p&=\frac{c}{n\sin\alpha},
\label{eq:wave}
\end{align}
where $k_0=\omega/c$, $n\sin\alpha>1$ ensures evanescence, and $\beta_p=v_p/c$. For the symmetric, in-phase carrier component,
\begin{align}
F_z&\simeq qE_m\cosh(\kappa x)\cos\phi,\label{eq:fz}\\
F_x&\simeq q\frac{k_z}{\kappa}E_m(1-\beta_z\beta_p)
\sinh(\kappa x)\sin\phi.\label{eq:fx}
\end{align}
Here $q=-e$, $E_m>0$, and $\phi=k_zz-\omega t+\phi_0$. In this convention, $\pi/2<\phi<\pi$ yields acceleration and focusing near the axis, while $\pi<\phi<3\pi/2$ yields acceleration and defocusing. These are local carrier-force intervals, not phases extracted from a bunch-averaged field. Net energy gain depends on integrated work rather than remaining in one accelerating half-cycle.

For a near-axis reference electron with $\beta=\sqrt{1-\gamma^{-2}}$,
\begin{align}
\frac{d\phi}{dz}&=k_0\left(n(z)\sin\alpha-\beta^{-1}\right),
\label{eq:phase}\\
\frac{d\gamma}{dz}&=\frac{qE_m(z)}{m_ec^2}\cos\phi.
\label{eq:energy}
\end{align}
Writing $\overline b_i=L_i^{-1}\int_0^{L_i}ds/\beta_i(s)$ for the section-averaged inverse velocity gives
\begin{align}
\Delta\phi_i&=k_0L_i(n_i\sin\alpha-\overline b_i),
\label{eq:segment_budget}\\
\Delta\phi_{\mathrm{tot}}&=k_0\left[\sin\alpha\sum_{i=1}^N n_iL_i-\int_0^L\frac{dz}{\beta(z)}\right].
\label{eq:global_budget}
\end{align}
Equivalently, with $\overline n_{\mathrm{syn},i}=\overline b_i/\sin\alpha$,
\begin{equation}
(n_i-\overline n_{\mathrm{syn},i})L_i
=\frac{\Delta\phi_i}{k_0\sin\alpha}.
\label{eq:nL_relation}
\end{equation}
For a specified section phase excursion, $n_i$ and $L_i$ are coupled through the electron energy evolution: the average inverse velocity depends on the energy gained in preceding sections and within the current section. This relation is implicit, not a fixed-index design formula. Section contributions of opposite sign can compensate for one another in the accumulated phase budget, although a small final phase shift is neither assumed for the numerical solution nor sufficient by itself to ensure net acceleration. In a constant-index section, acceleration introduces curvature into the phase trajectory, while a step in refractive index changes its slope:
\begin{align}
\phi''&=\frac{k_0\gamma'}{(\gamma^2-1)^{3/2}},
\label{eq:phase_curvature}\\
\left[\phi'\right]_{i}^{i+1}&=k_0\sin\alpha(n_{i+1}-n_i),
\label{eq:slope_reset}
\end{align}
where primes denote derivatives with respect to $z$. These relations assume that the particle velocity is continuous and that no additional carrier-phase jump occurs at an interface. The evanescent and synchronous indices are $n_{\mathrm{crit}}=1/\sin\alpha$ and $n_{\mathrm{syn}}=1/(\beta\sin\alpha)$. Negative slippage requires $n_{\mathrm{crit}}<n<n_{\mathrm{syn}}$, leaving a relativistic index interval
\begin{equation}
\Delta n_-=n_{\mathrm{syn}}-n_{\mathrm{crit}}
\simeq\frac{1}{2\gamma^2\sin\alpha}.
\label{eq:index_scaling}
\end{equation}
For a section in which the electron accelerates monotonically and accumulates negative slippage, the magnitude of the slip rate is bounded by $k_0(\beta_{i,\mathrm{in}}^{-1}-1)$. Consequently, a prescribed negative excursion requires
\begin{equation}
L_i>\frac{|\Delta\phi_i|}{k_0(\beta_{i,\mathrm{in}}^{-1}-1)}
\simeq\frac{2\gamma_{i,\mathrm{in}}^2|\Delta\phi_i|}{k_0}.
\label{eq:length_scaling}
\end{equation}
The index window narrows as $\gamma^{-2}$ while the required length grows as $\gamma^2$. This bound assumes no substantial deceleration and is a carrier-level design constraint, not a reconstructed particle phase in the broadband pulse.

At $\alpha=27.13^\circ$, the section lengths (in $\mu$m) are $(156,256,69,185,159,387,333)$ and the corresponding target effective refractive indices are $(4.253,2.806,2.647,2.257,2.639,2.219,2.704)$, giving $L=\SI{1545}{\micro\meter}$. The seven-section sequence is one numerical example of the phase-trajectory control framework. Possible routes to realizing tailored refractive indices include reported THz dielectrics, compositionally adjustable chalcogenide glasses, and subwavelength effective-index structures~\cite{31,32,33,34,35,36,37}. The first-section target $n_1\simeq4.253$ is particularly demanding; engineered high-index THz structures provide a precedent~\cite{38}, but this index sequence is not an experimentally validated material stack. The numerical design was selected through a coarse-to-fine parameter search that rejected high-gain trajectories exhibiting excessive transverse expansion. The refractive-index values are rounded numerical design parameters rather than validated fabrication tolerances. Material loss, dispersion, finite-width effects, and interface coupling require further assessment for practical implementation.

Electromagnetic fields calculated using CST Studio Suite~\cite{39} are imported into General Particle Tracer (GPT)~\cite{40} for particle tracking. The incident THz drive consists of two equal-amplitude, in-phase pulses represented by a tabulated near-single-cycle waveform at \SI{3}{THz}. The incident peak electric-field magnitude is set to \SI{1}{GV\per\meter}, and the rms width of a Gaussian fit to the field envelope is approximately \SI{0.095}{ps}. The injected electron bunch has an initial mean kinetic energy of \SI{100}{keV} and a charge of \SI{1}{\femto\coulomb}. It is represented by 5000 macroparticles whose initial transverse and longitudinal positions follow truncated Gaussian distributions. The initial normalized transverse emittance is \nmrad{1}, the relative rms kinetic-energy spread is \SI{0.1}{\percent}, and the injection Twiss parameters are $\alpha_{x,0}=3.5$ and $\beta_{x,0}^{\mathrm{Tw}}=\SI{0.26}{mm}$. The corresponding initial rms transverse beam size is approximately \SI{0.62}{\micro\meter}. The complete beam and THz-drive parameters are summarized in Table~\ref{tab:beam_thz}. GPT performs three-dimensional particle tracking, including mesh-based space-charge effects, while the applied driving field is assumed to be invariant along the out-of-plane direction~\cite{29}. The uniform-prism reference uses the same incident THz drive and injection conditions. Near-exit beam properties are evaluated from the particle distribution at the fixed time $t=\SI{6.12}{ps}$, when the bunch is near the nominal exit of the \SI{1545}{\micro\meter} structure.

\begin{table}[t]
\caption{Simulation parameters.}
\label{tab:beam_thz}
\begin{ruledtabular}
\begin{tabular}{ccc}
Symbol & Physical quantity & Value\\
\hline
$K_0$ & Initial mean kinetic energy & \SI{100}{keV}\\
$Q_b$ & Bunch charge & \SI{1}{\femto\coulomb}\\
$\epsilon_{n,x,0}$ & Initial normalized transverse emittance & \nmrad{1}\\
$r_{K,0}$ & Initial relative rms kinetic-energy spread & \SI{0.1}{\percent}\\
$\alpha_{x,0}$ & Initial Twiss parameter & 3.5\\
$\sigma_{x,0}$ & Initial rms transverse beam size & \SI{0.62}{\micro\meter}\\[3pt]
$f_{\mathrm{THz}}$ & THz frequency & \SI{3}{THz}\\
$\sigma_{t,\mathrm{env}}$ & RMS field-envelope width & \SI{0.095}{ps}\\
$E_{\mathrm{pk}}$ & Incident peak electric-field magnitude & \SI{1}{GV\per\meter}\\
\end{tabular}
\end{ruledtabular}
\end{table}

\begin{figure}[!t]
\centering
\includegraphics[width=0.98\columnwidth]{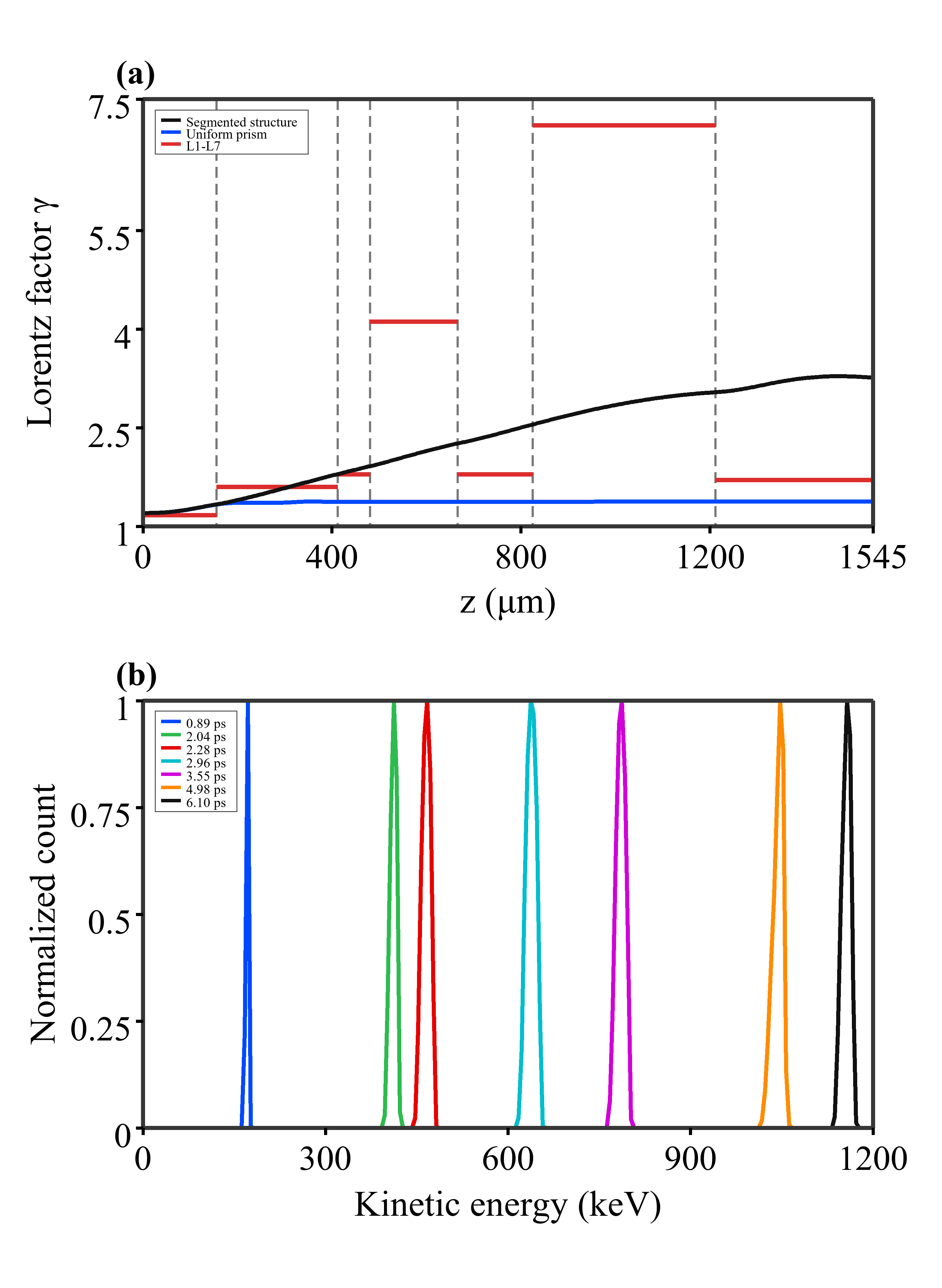}
\caption{Longitudinal beam dynamics. (a) Evolution of the bunch-averaged Lorentz factor in the segmented and uniform-prism structures. The red horizontal segments represent the phase Lorentz factors $\gamma_{p,i}$ of the seven sections, and the vertical dashed lines indicate the section boundaries. (b) Normalized kinetic-energy spectra of the segmented bunch at selected times during acceleration.}
\label{fig:energy}
\end{figure}

Fig.~\ref{fig:energy}(a) compares the evolution of the bunch-averaged Lorentz factor in the segmented accelerator and the uniform-prism reference. The segmented bunch exhibits sustained energy gain, whereas the uniform-prism reference shows an initial increase followed by saturation. The red horizontal segments indicate the phase Lorentz factors $\gamma_{p,i}$ of the individual sections. By adjusting the local phase velocity through the refractive-index sequence, the phase-velocity mismatch and accumulated phase slippage can be controlled along the accelerator, as described by Eq.~\eqref{eq:phase}. Fig.~\ref{fig:energy}(b) presents the normalized kinetic-energy spectra at selected times. The successive spectra shift toward higher kinetic energies, illustrating the progressive acceleration of the electron bunch into the MeV energy range.

\begin{figure}[!t]
\centering
\includegraphics[width=0.98\columnwidth]{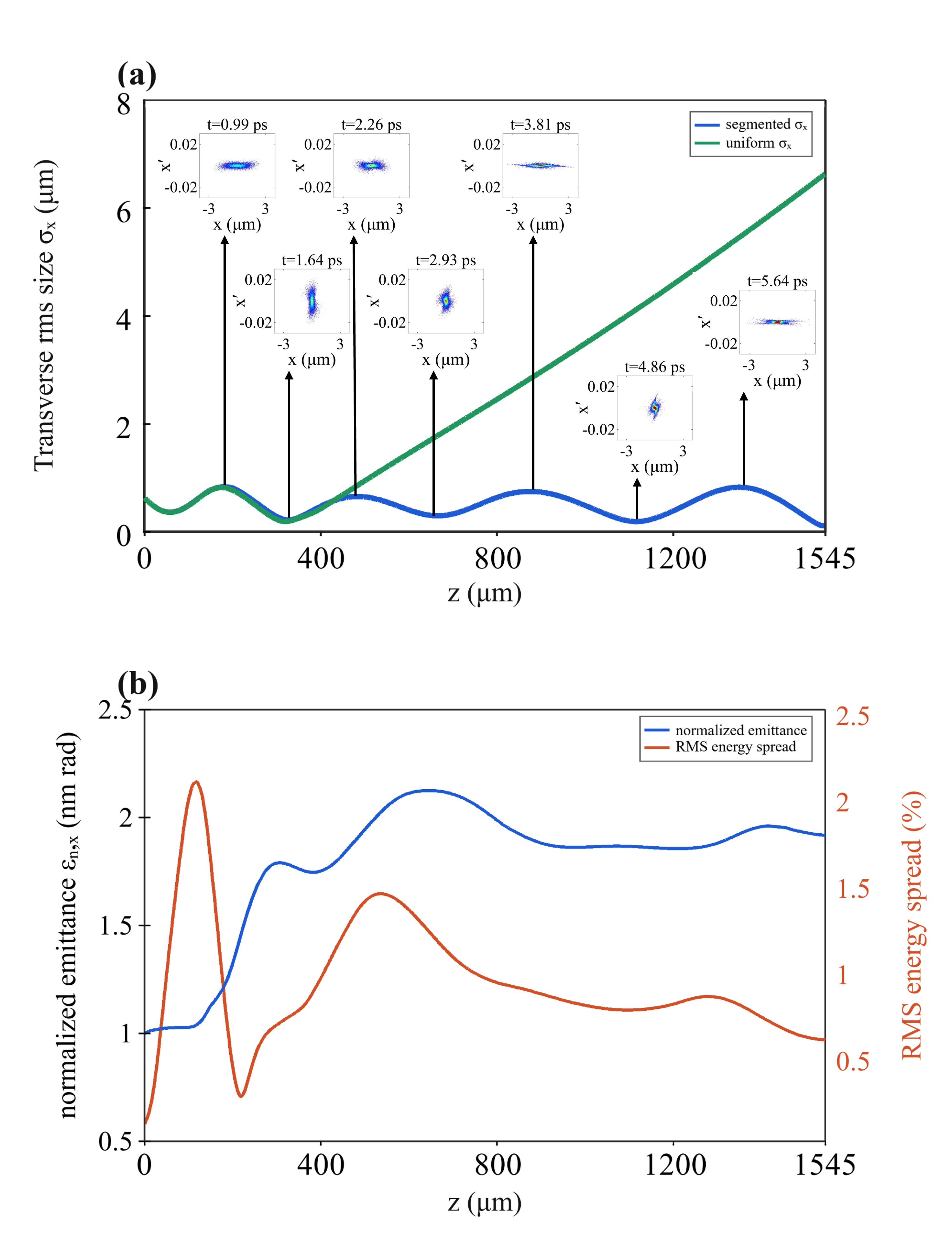}
\caption{Transverse confinement and beam quality. (a) Rms transverse beam size in the segmented and uniform-prism structures; the insets show the transverse $(x,x^{\prime})$ phase space at selected times, obtained from a separate 10,000-particle simulation. The main curves use 5000 macroparticles. (b) Evolution of the normalized rms transverse emittance (blue) and relative rms kinetic-energy spread (orange).}
\label{fig:transverse}
\end{figure}

The near-exit mean Lorentz factor is $\langle\gamma\rangle\simeq3.2597$. The corresponding mean kinetic energy is \SI{1154.7}{keV}, giving an energy gain of \SI{1054.7}{keV}. Over \SI{1.545}{mm}, the corresponding effective accelerating gradient is \SI{682.7}{MeV\per\meter}. The near-exit relative rms kinetic-energy spread is approximately \SI{0.6}{\percent}, as shown in Fig.~\ref{fig:transverse}(b). At the same sampling time, the uniform-prism reference reaches a mean kinetic energy of approximately \SI{191}{keV}, corresponding to an energy gain of about \SI{91}{keV} relative to the initial kinetic energy of \SI{100.0}{keV}.

\begin{figure*}[!t]
\centering
\includegraphics[width=0.98\textwidth]{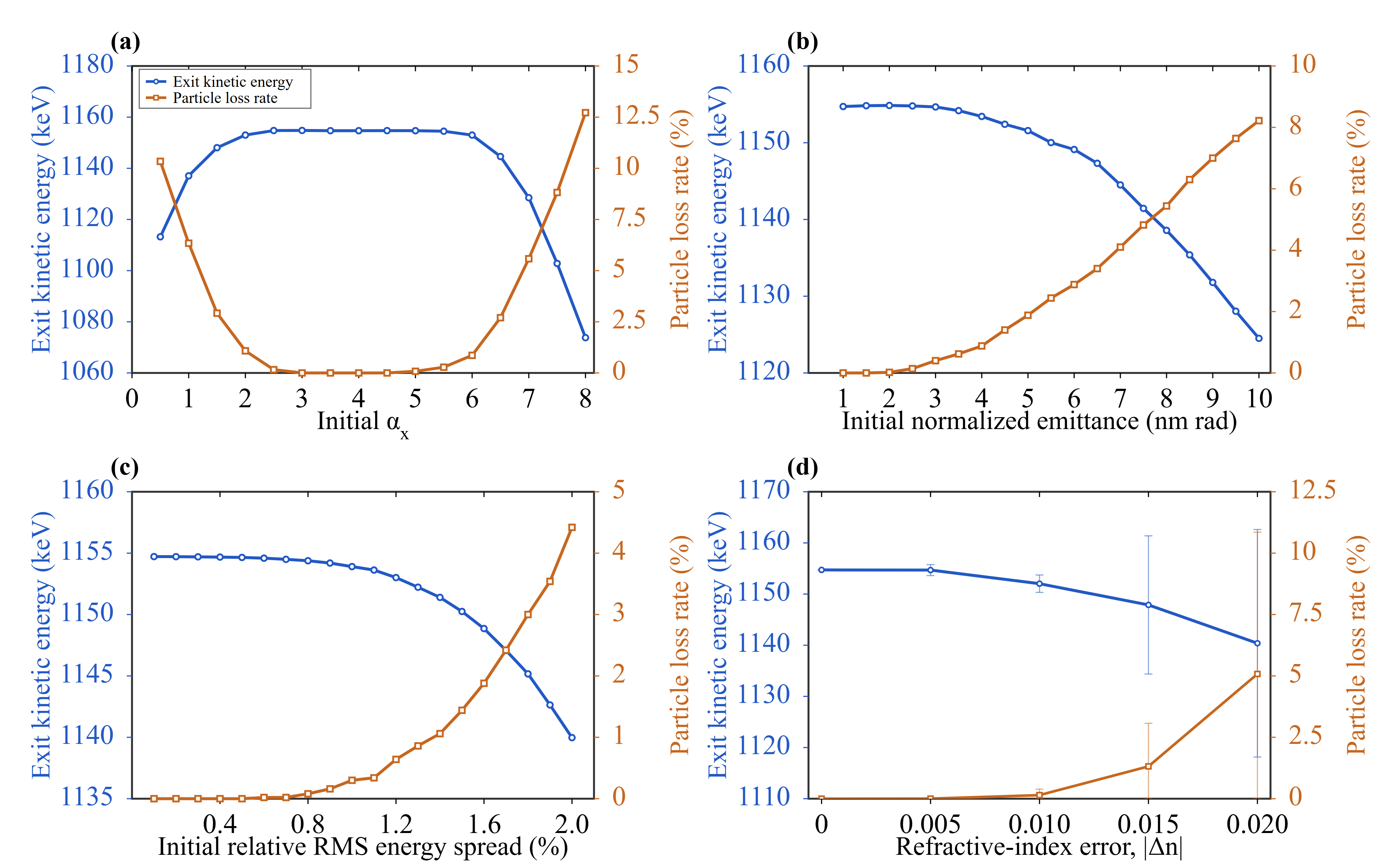}
\caption{Sensitivity of the exit mean kinetic energy (blue) and reported particle-loss rate (orange) to (a) the initial Twiss parameter $\alpha_{x,0}$, (b) the initial normalized rms transverse emittance, (c) the initial relative rms kinetic-energy spread, and (d) the absolute refractive-index perturbation amplitude $|\Delta n|$. The error bars in (d) represent sample standard deviations obtained from independent refractive-index perturbation realizations.}
\label{fig:sensitivity}
\end{figure*}

The segmented bunch undergoes repeated transverse-envelope compression and expansion, as shown in Fig.~\ref{fig:transverse}(a). The transverse rms beam size remains below \SI{1}{\micro\meter} throughout the segmented structure, well within the \SI{5}{\micro\meter} channel half-width. In contrast, the rms beam envelope in the uniform-prism reference expands beyond the channel half-width. The transverse $(x,x^{\prime})$ phase-space distributions exhibit successive rotation and shear during propagation, consistent with the repeated compression and expansion of the beam envelope. These results indicate that the proposed phase-trajectory control strategy, combined with suitable injection conditions, provides effective transverse rms beam confinement.

Fig.~\ref{fig:transverse}(b) shows the evolution of the normalized rms transverse emittance and relative rms kinetic-energy spread. The normalized transverse emittance increases from \nmrad{1} to approximately \nmrad{1.92}, with oscillations during acceleration. The relative rms kinetic-energy spread exhibits pronounced transient growth, exceeding \SI{2.00}{\percent} before decreasing to approximately \SI{0.60}{\percent} near the exit. These results demonstrate that a compact transverse beam envelope can be maintained despite finite emittance growth and non-monotonic energy-spread evolution.

Fig.~\ref{fig:sensitivity} evaluates the sensitivity of the exit mean kinetic energy and particle retention to variations in the injection and structural parameters. Panels (a)--(d) examine the effects of the initial Twiss parameter $\alpha_{x,0}$, the initial normalized transverse emittance, the initial relative rms kinetic-energy spread, and the absolute refractive-index perturbation amplitude $|\Delta n|$, respectively. In the refractive-index scan, the refractive indices of the seven sections are perturbed independently in each realization, $n_i^{(r)}=n_i+\delta n_i^{(r)}$, where $i=1,\ldots,7$ labels a section and $r$ labels an independent realization. The error amplitude $|\Delta n|$ characterizes the prescribed magnitude of these perturbations. The electromagnetic fields of each perturbed configuration are recalculated in CST before GPT tracking; other nominal settings are held fixed.

The Twiss-parameter scan in Fig.~\ref{fig:sensitivity}(a) shows that the exit mean kinetic energy increases from \SI{1113.2}{keV} at $\alpha_{x,0}=0.5$ to \SI{1154.7}{keV} at the nominal value of $\alpha_{x,0}=3.5$, while the reported particle-loss rate decreases from \SI{10.34}{\percent} to zero. At $\alpha_{x,0}=3.0$, the mean kinetic energy reaches \SI{1154.8}{keV}, also with zero reported particle loss. Increasing $\alpha_{x,0}$ to 8.0 reduces the mean kinetic energy to \SI{1073.8}{keV} and increases the loss rate to \SI{12.72}{\percent}. These results identify a favorable injection range in which high mean kinetic energy and low particle loss can be simultaneously achieved. Fig.~\ref{fig:sensitivity}(b) shows that increasing the initial normalized transverse emittance from \nmrad{1} to \nmrad{10} reduces the mean kinetic energy from \SI{1154.7}{keV} to \SI{1124.5}{keV}, while the particle-loss rate increases from zero to \SI{8.22}{\percent}. At \nmrad{2}, zero reported particle loss is retained. In the initial energy-spread scan shown in Fig.~\ref{fig:sensitivity}(c), increasing the relative rms kinetic-energy spread from \SI{0.1}{\percent} to \SI{1}{\percent} yields an exit mean kinetic energy of \SI{1153.9}{keV} and a particle-loss rate of \SI{0.3}{\percent}. At \SI{2}{\percent}, these values become \SI{1140}{keV} and \SI{4.42}{\percent}, respectively. The injection-parameter scans demonstrate that the segmented structure maintains relatively high energy gain over a range of initial beam conditions, although increasing emittance and energy spread lead to greater particle loss.

Figure~\ref{fig:sensitivity}(d) presents the sensitivity of the acceleration performance to refractive-index perturbations. For each perturbation amplitude, ten independent random configurations of the refractive indices of the seven sections are generated. At $|\Delta n|=0.01$, the mean kinetic energy is \SI{1152.0}{keV}, with a mean particle-loss rate of \SI{0.15}{\percent}. When the perturbation amplitude increases to $|\Delta n|=0.02$, the mean kinetic energy decreases to \SI{1140.4}{keV}, while the mean particle-loss rate increases to \SI{5.08}{\percent}. At this amplitude, the standard deviations across the ten realizations are \SI{22.2}{keV} in the run-mean kinetic energy and 5.78 percentage points in the particle-loss rate. Because the accumulated phase depends on the product of $n_i$ and $L_i$ rather than on $n_i$ alone, a realized index deviating by a known $\delta n_i$ can be partially compensated by retuning the corresponding section length to restore the target $\Delta \phi_i$, provided the transverse field profile remains approximately unchanged. The loss fractions in Fig. 4(d) are therefore a conservative lower bound: they characterize fixed-length sensitivity, whereas allowing length rescaling would broaden the practical index tolerance. Moreover, the trend in Fig. 4(d) indicates that the baseline refractive-index combination is not yet saturated and leaves room for further refinement toward improved performance.

In conclusion, sustained inverse-Cherenkov acceleration with transverse confinement can be achieved not by freezing the transverse phase space, but by coordinating phase-trajectory design with injection matching. The section-wise refractive-index and length sequences govern accumulated phase slippage and the evolution of the beam envelope, while the injected Twiss parameters determine how the initial phase-space distribution evolves along this trajectory. In a seven-section numerical demonstration, a 100-keV electron bunch reaches a mean kinetic energy of \SI{1154.7}{keV} over \SI{1.545}{mm}, corresponding to an energy gain of \SI{1054.7}{keV} and an effective accelerating gradient of \SI{682.7}{\mega\volt\per\meter}. The bunch maintains a submicrometer transverse rms size, with an exit normalized emittance of \nmrad{1.92} and a relative RMS energy spread of \SI{0.6}{\percent}. Crucially, the admissible injector emittance extends to at least \nmrad{2.0} without particle loss under the specified tracking-acceptance criterion---approximately two orders of magnitude above the $20\,\mathrm{pm}\text{-}\mathrm{rad}$ near-cold baseline considered here. Since the beam undergoes predominantly drift-like motion along the out-of-plane $y$ direction, we focus our discussion on the in-plane $x$ dynamics. Nevertheless, the $y$ coordinate is fully included in the three-dimensional particle tracking, which confirms that the bunch also remains well contained within the channel along this direction. These results establish coordinated phase-trajectory control and injection matching as a design strategy for relaxing injector-quality constraints in compact dielectric acceleration, shifting the emphasis from finding a sufficiently cold beam to matching the beam to a deliberately designed trajectory.

\begin{acknowledgments}
This work was supported by the National Natural Science Foundation of China (12125508, 12541503), the National Key Research and Development Program of China (2024YFA1612104), and the International Cooperation Program, Shanghai Municipal Commission of Science and Technology (26140780400). We also thank Zipeng Liu and Bangjie Deng for their valuable assistance and helpful discussions.
\end{acknowledgments}

\bibliographystyle{apsrev4-2}
\bibliography{reference}

\end{document}